# Assessing AI-Enhanced Single-Sweep Approximations for Problems with Forward-Peaked Scattering in Slab Geometry

Japan K. Patel[1,*], Matthew C. Schmidt[2], Anthony Magliari[3], and Todd A. Wareing[3]

[1]The Ohio State University, Columbus, OH; [2]Washington University in St. Louis, St. Louis, MO; [3]Varian Medical Systems, Palo Alto, CA

## ABSTRACT

While the Boltzmann transport equation can accurately model transport problems with highly forward-peaked scattering, obtaining its solution can become arbitrarily slow due to near-unity spectral radius associated with source iteration. Standard acceleration techniques like diffusion synthetic acceleration and nonlinear diffusion acceleration obtain merely one order of magnitude speedups compared to source iteration due to slowly decaying error moments. Additionally, converging approximations to the Boltzmann equation like Fokker-Planck and Boltzmann Fokker Planck run into similar problems with slow convergence. In this paper we assess the feasibility of using Fourier neural operators to obtain AI-enhanced low order, and single-sweep solutions for the transport equation in slab geometry using a predictor-corrector framework.

## 1. INTRODUCTION

Forward-peaked transport finds application in several important research areas including radiation oncology physics, plasma physics, astrophysics, shielding, etc. Such problems are characterized by small mean free paths with nearly singular differential scattering cross-sections in the forward direction and require relatively large expansions to represent scattering adequately. Several approximations, including Fokker-Planck (FP) [1], generalized Fokker Planck [2], Boltzmann Fokker Planck (BFP) [3], and generalized Boltzmann Fokker Planck (GBFP) [4] have been developed to address this problem by manipulating the scattering term of the Boltzmann transport equation (BTE). Additionally, acceleration techniques like diffusion synthetic acceleration (DSA) [5], nonlinear diffusion acceleration (NDA) [6], transport synthetic acceleration (TSA) [7], Angular multigrid (AMG) [8], Fokker Planck synthetic acceleration (FPSA) [9], nonlinear and modified Fokker Planck acceleration (MFPA) [10], modified $P_L$ acceleration (MPA) [11], quasi diffusion [12], etc. have also been investigated to speed up convergence for forward-peaked transport problems. Standard methods like DSA and NDA do not speed up calculations as effectively as they do for isotropic and weakly anisotropic problems because these methods do not attenuate higher order error moments [9]. Other methods like FPSA and AMG that attenuate higher order error moments have been demonstrated to be effective in 1D, but such methods are highly problem dependent [8,9]. To our knowledge, no single acceleration or approximation approach has been shown to work flawlessly in all transport regimes. In this paper, we explore using neural operators to approximate converged solutions without having to go through an exhaustive iteration process. Preliminary studies involving use of Fourier neural operator neural network (FNO) to approximate high order fine mesh transport solutions has been summarized in [13, 14]. In this paper, we investigate predictor-corrector frameworks to estimate higher fidelity solutions using lower order approximations. Predictors employed in this paper include $S_2$ discrete ordinates solutions, along with single-sweep estimates associated with source iteration, and pure absorbers. To our knowledge, this is the first time operator learning is being explored to develop single-sweep approximations to transport solutions.

---

*patel.3545@osu.edu

We primarily work with the slab geometry transport equation and begin by reviewing it. With angular flux $\psi$, $l^{th}$ Legendre moment of angular flux $\phi_l$, total macroscopic cross-section $\sigma_t$, $l^{th}$ order Legendre polynomial $P_l$, scattering cross-section moments $\sigma_{s,l}$, and an internal source $Q$, we consider the following slab geometry transport equation [15]:

$$\mu \frac{d}{dx}\psi(x,\mu) + \sigma_t(x)\psi(x,\mu) = \sum_{l=0}^{L} \frac{(2l+1)}{2} P_l(\mu)\sigma_{s,l}(x)\phi_l(x) + Q(x,\mu). \tag{1}$$

with moments of the angular flux are determined using the following relation:

$$\phi_l(x) = \int_{1}^{1} \psi(x,\mu)P_l(\mu)d\mu. \tag{2}$$

The following sections will introduce the predictor-corrector approach. We provide a brief overview of FNO the next section, along with some notes on how to use them to speed up transport calculations and conclude in Section 3.

## 2. NEURAL OPERATOR BASED APPROXIMATIONS

The main goal of operator learning is to find properties of systems and equations from data [16]. Traditionally, neural networks have generally been used to describe mappings from one vector to another. This is limiting because these mappings lack discrete-continuous consistency. Recent neural operator architectures overcome this drawback by inferring operators that map input to output functions. Several different types of neural operator architectures have been explored in the literature [16, 17]. In this paper, we focus on the FNO framework [17].

### 2.1. Fourier Neural Operator Neural Network

While a detailed explanation of an FNO can be found in [17] we briefly outline noteworthy details and implementation steps next. With two functions $a$ and $u$ of independent variables $x$ on domain $D$, and a mapping $G$ such that

$$u(x) = G(a(x)), \tag{3}$$

the goal is to determine an approximation $G_\theta \approx G$ with learnable parameters $\theta$, where,

$$G_\theta = G_0 \circ G_1 ... \circ G_K, \tag{4}$$

is a neural network composed of network layers $G_k$ represented according to

$$G_k v(x) = \sigma\left(B_k + \int_D K(x,y)v(y)dy\right). \tag{5}$$

Here, $\sigma$, $B_k$, $K(x,y)$, and $v(x)$ represent the activation function, bias function, convolution kernel, and an intermediate network function. With $N$ neurons per layer, this representation requires roughly $O(N^2)$ operations for evaluation. Li et al. [17] introduced the idea of using Fourier transforms to convert the convolution into a simple product in Fourier space and reduced the operational overhead to $O(NlogN)$. FNOs consist of four main components – the lifting layer, Fourier layers, nonlinear activation, and projection layer. The lifting layer projects input functions into a higher dimensional feature space by linear transformation $P$

$$v_0(x) = P(x). \tag{5}$$

Fourier layers apply Fourier transforms according to Eq. 6a, then multiply by learnable weights $W_\theta$ in the Fourier space as shown in Eq. 6b, and apply inverse Fourier transforms and add in bias and residual connections according to Eq. 6c [17].

$$\hat{v}_j(k) = F(v_j)(k). \tag{6a}$$

$$\hat{v}_j'(k) = W_\theta \hat{v}_j(k) \tag{6b}$$

$$v_{j+1}(x) = F^{-1}(\hat{v}_j')(x) + v_j(x) \tag{6c}$$

Further, a nonlinearity is introduced by pointwise applying a nonlinear activation function $\sigma$ according to

$$v_{j+1}(x) = \sigma\left(v_{j+1}(x)\right). \tag{6d}$$

The solution is obtained by projecting output onto the function space using projection operator $Q$

$$u(x) = Q(v_K)(x). \tag{7}$$

A pictorial representation of this setup is presented in Fig. 1 [17]

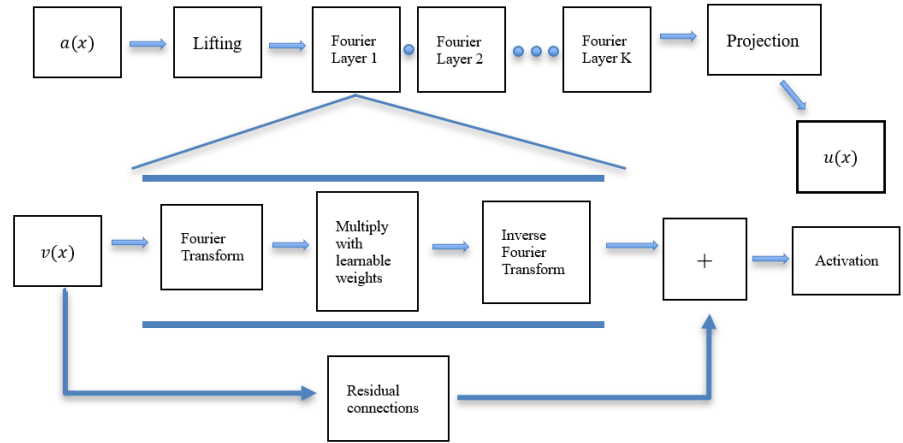

**Figure 1. Fourier Neural Operator Neural Network Framework.**

Learnable parameters such as lifting, projection, weighting, and bias functions can be inferred by minimizing the loss function

$$\epsilon_\theta = \frac{1}{N}\sum_i^N \left\|u_{i,FNO} - u_i\right\|^2, \qquad (8)$$

using standard optimizers [18].

## 2.2. Predictor-Corrector Approximation Framework

We employ standard predictor-corrector setup to refine lower fidelity solutions to obtain higher fidelity estimates. In this section we consider three types of predictors – single sweep $S_N$, pure absorber, and fully converged $S_2$ solutions. The predicted solutions are then passed on to the FNO architecture to implicitly correct (map) predicted approximations to higher fidelity estimates. We begin by introducing some preliminaries. Rewriting Eq. 1 in operator notation,

$$L\psi(x,\mu) = S\psi(x,\mu) + Q(x,\mu), \qquad (9)$$

with

$$L = \mu\frac{d}{dx} + \sigma_t(x), \qquad (10a)$$

$$S = \sum_{l=0}^{L} \frac{(2l+1)}{2} P_l(\mu)\sigma_{s,l}(x)\phi_l(x), \qquad (10b)$$

the discrete ordinates formulation with angle index $n$ takes the following form [13]

$$L\psi(x,\mu_n) = S\psi(x,\mu_n) + Q(x,\mu_n). \qquad (11)$$

With iteration index **m**, source iteration for $S_N$ equations takes the following form

$$L\psi^{m+1}(x,\mu_n) = S\psi^m(x,\mu_n) + Q(x,\mu_n). \qquad (12)$$

The predictor-corrector approximation procedure breaks solution into two parts. Obtaining a low order (LO) approximation, and correcting it to approximate higher order (HO) solution:

$$\text{Predict: } \phi_{0,LO}, \qquad (13a)$$

$$\text{Correct: } G(\phi_{0,LO}) \approx \phi_{0,HO} \qquad (13b)$$

This is similar to the predict-correct-iterate framework [9] employed for synthetic acceleration, except the iterate part is omitted. We train FNOs with 4 Fourier layers containing 64 Fourier modes each, along with the ReLU activation function [19] using the JAX framework [20]. Each case has four in-channels and one out-channel. While the input function varies depending on the choice of the function being mapped to the higher fidelity solution, each scenario additionally takes in the anisotropy factor and scattering ratio as inputs. The ADAM optimizer was used to minimize training losses [21] with learning rate of $1\times10^{-4}$ and batch size of 25. An average relative $L_2$ error norm was used to characterize the quality of inferred mappings.

## FNO-enhanced single sweep solutions

The first predictor approach considered in this paper is the scalar flux obtained after one source iteration step (SS):

$$\psi^1(x,\mu_n) = L^{-1}S\psi^0(x,\mu_n) + L^{-1}Q(x,\mu_n), \tag{14a}$$

with higher fidelity approximation obtained using FNO such that

$$\phi_{0,HO} \approx G_{\theta,SS}(\phi_0^1). \tag{14b}$$

## FNO-enhanced pure absorber solutions

The second predictor method assumes a pure absorber (PA) system and ignores all scatter. The solution to such problems converges in one iteration and therefore this approach can also be deemed "single-sweep". In this scenario, Eq. 11 takes the following form

$$\psi(x,\mu_n) = L^{-1}Q(x,\mu_n). \tag{15a}$$

The predictor $\phi_{0,PA}$ obtained by solving Eq. 15a, and the correction is obtained by mapping this solution to higher fidelity solution using an FNO:

$$\phi_{0,HO} \approx G_{\theta,PA}(\phi_{0,PA}). \tag{15b}$$

We note that this approach becomes a special case of the SS approach when initial flux is set to zero in Eq. 14a. This, however, is not the case here as we initialize the flux according to the beam source. Therefore, the two approaches return slightly different input fluxes.

## FNO-enhanced S$_2$ solutions

The final predictor approach considered in this paper converges the S$_2$ solutions to desired tolerance according to

$$L\psi^{m+1}(x,\mu_1) = S\psi^m(x,\mu_1) + Q(x,\mu_1), \tag{16a}$$

$$L\psi^{m+1}(x,\mu_2) = S\psi^m(x,\mu_2) + Q(x,\mu_2), \tag{16b}$$

to obtain solution $\phi_{0,S_2}$ and maps it back to the HO solution using an FNO such that

$$\phi_{0,HO} \approx G_{\theta,S_2}(\phi_{0,S_2}). \tag{16c}$$

Note that while the method to train FNOs $G_{\theta,SS}$, $G_{\theta,PA}$, and $G_{\theta,S_2}$ is the same, these are three distinct mappings. In other words, each predictor has a companion corrector mapping, and they must always be used in conjunction with each other.

### 2.3. Numerical Experiments

For preliminary assessment of FNO-enhanced solutions, we consider a slab geometry problem of size 50 cm with a beam source along the most forward direction. We incorporate anisotropy into the problem using the Henyey-Greenstein kernel, whose scattering moments are evaluated according to [22]

$$\sigma_{s,l}^{HGK} = \sigma_s g^l. \tag{17}$$

As the anisotropy factor $g$ goes from zero to unity, scattering goes from isotropic to highly anisotropic and forward peaked. FNO correctors are trained using a dataset of 512 cases with random material properties. The transport equation is discretized using linear discontinuous (LD) finite element method and solved using the source iteration algorithm on a Precision 7920 workstation with two Intel Xeon Gold 6148 CPUs and 192 GB of RAM. While $S_{16}P_{15}$ estimates represent HO solutions, LO solutions are obtained using predictor setups described in Sec. 2.2. The anisotropy factor for the Henyey-Greenstein phase function is randomly varied from 0.8 to 1 and the dataset is divided into training and test sets in 3:1 proportion. An NVIDIA A100 GPU was used via Google Colab for training which took roughly 4 minutes for each of the predictor-corrector scenario. It is noteworthy that changing to an NVIDIA T4 GPU increased training runtime to roughly 14 minutes. Training and inference were found to be hardware dependent as expected. Figures 2, 3, and 4 present example cases for the three predictor-corrector setups. The scattering ratio and anisotropy factors for this case are 0.995 and 0.977, respectively. Similarly, a general runtime comparison is not possible because spectral radius is highly problem dependent.

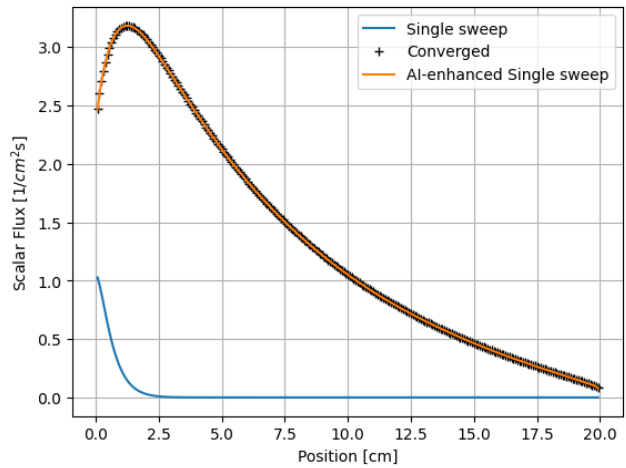

**Figure 2. Scalar flux comparison with SS predictor.**

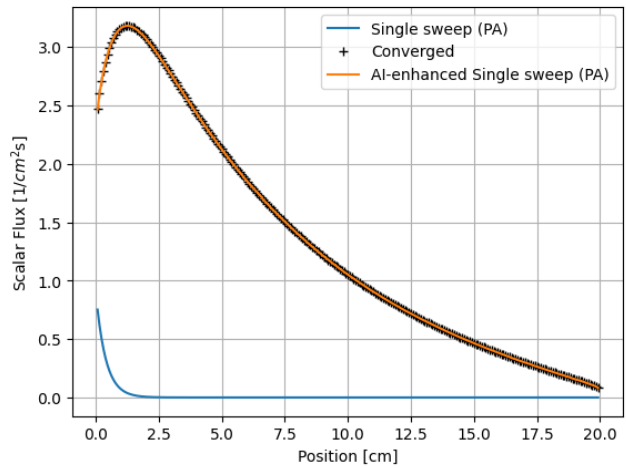

**Figure 3. Scalar flux comparison with PA predictor.**

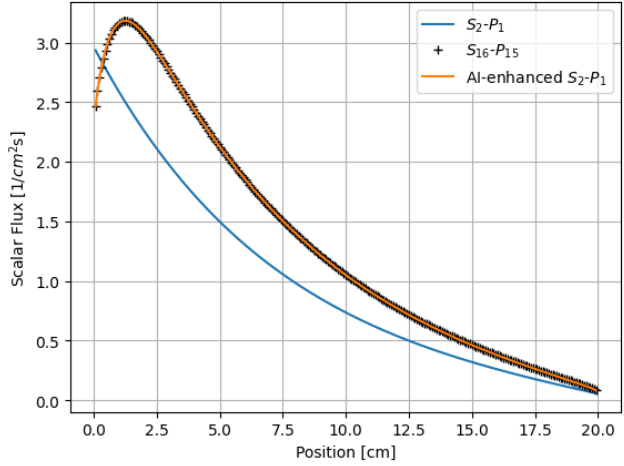

Figure 4. Scalar flux comparison with $S_2$ predictor.

Table I. Average Speedups and Errors.

| Predictor | SS | PA | Converged $S_2$ |
|---|---|---|---|
| $\dfrac{\bar{t}_p}{\bar{t}_{HO}}$ | $6.12 \times 10^{-2}$ | $3.66 \times 10^{-2}$ | $1.37 \times 10^{-2}$ |
| $\bar{t}_c$ [s] | $4.80 \times 10^{-4}$ | $4.46 \times 10^{-4}$ | $4.53 \times 10^{-4}$ |
| $\bar{\epsilon}$ | $1.27 \times 10^{-3}$ | $1.31 \times 10^{-3}$ | $1.11 \times 10^{-3}$ |

We observe from Figures 2, 3, and 4 that while stand-alone LO and single-sweep approximations fail to capture the physics, incorporating FNO-based corrections improves solution quality significantly for all three predictors. Our general observation was that if appropriate physics parameters are provided as input, and low order approximations capture overall variability in solutions, FNOs work quite well. Having said that, we also observed that not using physical parameters with input, degrades mapping for SS because scattering ratio and anisotropy factors often play a crucial role in determining how the first sweep relates to the converged solution.

With average time to generate predictor estimate $\bar{t}_p$, that to obtain higher fidelity solution $\bar{t}_{HO}$, average time to obtain FNO-based corrections $\bar{t}_c$ and average relative $L_2$ error norms $\bar{\epsilon}$, we summarize average speedups and errors in Table 1. We observe, as expected, that FNO-based corrections take a fraction of the time required by source iteration to return higher fidelity solutions. Specifically, for the dataset used in this paper, obtaining FNO-based corrections was observed to be roughly 3-4 orders of magnitude faster than obtaining higher fidelity solutions by source iteration. Additionally, this is significantly larger than the two orders of magnitude speedup seen with evaluation of LO and single-sweep predictors. Therefore, we note that runtime speedups are dominated by the choice of LO predictors. For the cases under consideration, while we observe that converged $S_2$ solutions was the most efficient, the single-sweep methods roughly provided the same order of speedups. This pattern is not unexpected because $S_2$ solutions, with their coarse angular mesh converge relatively quickly and do not require as many iterations as higher fidelity solutions. Additionally, with their coarse mesh, computational overhead is also relatively smaller than that required while sweeping through finer meshes.

# 3. CONCLUSIONS

FNO-enhanced predictor-corrector setup was investigated using a small dataset of 512 samples. Three distinct predictor mechanisms were tested with all three methods proving to be sufficiently accurate and comparable in efficiency. In conclusion, we demonstrate FNO as a novel method for mapping lower fidelity predictions to higher fidelity solutions efficiently. Future work will focus on moving to full 3D transport solutions and investigating how to leverage FNO for preconditioning.

## ACKNOWLEDGMENTS

We would like to thank Prof. John C. Lee from the University of Michigan for discussions and Prof. Dean Wang from The Ohio State University for his kind support.